**Does Open Access Foster Interdisciplinary Citation? Decomposing Open Access Citation Advantage**

Kai Nishikawa[1, 2]

knishikawa@slis.tsukuba.ac.jp

0000-0001-5949-3610

Akiyoshi Murakami[2]

[1] Institute of Library, Information and Media Science, University of Tsukuba, 1-2, Kasuga, Tsukuba, 305-8550, Ibaraki, Japan.

[2] National Institute of Science and Technology Policy (NISTEP), Ministry of Culture, Science and Sports (MEXT), 3-2-2, Kasumigaseki, Chiyoda-ku, 100-0013, Tokyo, Japan.

**Abstract**

The existence of an open access (OA) citation advantage—that is, whether OA increases citations—has been a topic of interest for many years. Although numerous previous studies have focused on whether OA increases citations, expectations for OA go beyond that. One such expectation is the promotion of knowledge transfer across various fields. This study aimed to clarify whether OA, especially gold OA, increases interdisciplinary citations in various natural science fields. Specifically, we measured the effect of OA on interdisciplinary and within-discipline citation counts by decomposing an existing metric of the OA citation advantage. The results revealed that OA increases both interdisciplinary and within-discipline citations in many fields and increases only interdisciplinary citations in chemistry, computer science, and clinical medicine. Among these fields, clinical medicine tends to obtain more interdisciplinary citations without being influenced by specific journals or papers. The findings indicate that OA fosters knowledge transfer to different fields, which extends our understanding of its effects.

**Statements and Declarations**

*Competing Interests* The authors have no competing interests to declare that are relevant to the content of this article.

*Funding* No funding was received for conducting this study.

# Introduction

Since the release of the Budapest Open Access Initiative, the number of open access (OA) papers has grown steadily. With several countries' policies now mandating that publicly funded research outputs be made OA, the number of OA papers is expected to continue increasing in the future. A paper can be made OA in several ways. The most frequently selected method is gold OA, which entails publishing a paper in a fully OA journal in which all papers are published as OA (Heidbach et al., 2022). Gold OA requires authors to pay the publisher a fee for OA, called the article processing charge (APC), and countries that mandate OA often subsidize the APC[1].

In this policy context, the effects of OA are often evaluated through the lens of the OA citation advantage, which refers to the potential increase in the number of citations that OA publications receive compared to non-OA publications. Although research on this advantage has been ongoing since Lawrence (2001) proposed its existence, previous studies have produced differing results, and no generalizable finding has been established to indicate that the OA citation advantage exists consistently across all groups of papers (Langham-Putrow et al., 2021).

Additionally, most previous studies have focused on determining whether an OA citation advantage exists. Even when results suggest that OA increases citation counts, few studies have examined the types of citations in detail. However, expectations surrounding OA—and open science more broadly—should not be confined to boosting citation counts. One key expectation is that openness will foster knowledge transfer across fields and sectors (Organization for Economic Co-operation and Development [OECD], 2015).

Therefore, this study aims to clarify the effects of OA, particularly gold OA, on knowledge transfer across fields. Specifically, we decompose the traditional metric of the OA citation advantage by distinguishing between interdisciplinary and within-discipline citations and introduce a new metric to determine whether OA increases interdisciplinary citations. Using this metric, we explore whether papers across various natural science fields are more likely to be cited in papers from other fields when published as OA.

By focusing on interdisciplinary citations, this study introduces a novel perspective to previous OA citation advantage studies. The findings demonstrate how OA may enhance knowledge transfer across fields and provide OA practitioners, including policymakers, research funders, university libraries, and researchers, with evidence of the significance of their activities.

In the following section, we review related studies to identify the existing knowledge gap in OA and where this study stands. The methods used in this study are then described in detail, followed by the presentation and discussion of the results. Finally, the conclusions are presented, and future research directions are provided.

---

[1] If diamond OA is considered a subtype of gold OA, then paying an APC is not always required for gold OA.

## Literature review

**Open Access citation advantage**

Since Lawrence (2001) proposed the existence of the OA citation advantage in the field of computer science, numerous studies have sought to verify this phenomenon across various fields. Langham-Putrow et al. (2021) recently identified 134 relevant publications. However, it noted that researchers have yet to reach a consensus on whether an OA citation advantage exists. Although many previous studies encompassing multiple fields have reported evidence of an OA citation advantage in some fields, they concluded that a generalizable advantage is lacking (e.g., Dorta-Gonzalez et al., 2017; Dorta-Gonzalez & Santana-Jiménez, 2018). Langham-Putrow et al. (2021) attributed inconsistencies in the findings to several factors, including the different types of OA being examined and the absence of standardization in the definitions, metrics, and methodologies used to measure the OA citation advantage.

Early studies on the OA citation advantage proposed three postulates as mechanisms by which OA leads to more citations (Kurtz et al., 2005; Craig et al., 2007; Moed 2007; Davis et al., 2008). The first is the *open access postulate*, which suggests that OA papers are read and cited more frequently because of their greater availability. The second is the *early view postulate*, which suggests that OA papers tend to have higher cumulative citation counts because they are made available online earlier than other publications. The third, the *selection bias postulate*, states that papers authored by reputable authors or those of higher quality are more likely to be made OA, leading to increased citation counts for OA papers.

It has been suggested that a pure OA citation advantage corresponds to the first postulate and that the influences of the second and third postulates should be controlled to ascertain whether an OA citation advantage exists (Gaulé & Maystre, 2011; Niyazov et al., 2016; Dorta-Gonzalez et al., 2017). To control for the *early view postulate*, researchers often assess the number of citations of a paper within a specific citation time window (e.g., three years post-publication; Sotudeh et al., 2015; Dorta-Gonzalez et al., 2017). Additionally, Sotudeh et al. (2015) mentioned this postulate as particularly relevant for green OA, possibly because preprints in green OA are often made available without embargoes, allowing earlier access compared to version of records. To address the *selection bias postulate*, a common approach involves comparing OA and non-OA papers by categorizing them based on their journal impact factor (JIF) quartiles in their respective fields (Niyazov et al., 2016).[2]

Many studies have employed the mean or median citations per paper as metrics to measure the OA citation advantage (Langham-Putrow et al., 2021). Additionally, various other metrics have been utilized depending on the research purpose and context. In particular, studies investigating the OA citation advantage across fields often use the ratio of the difference between the average citation count of OA papers and non-OA papers to the average citation count of non-OA papers (as described in the Methods section, this metric is called the OACA in the present study; Sotudeh et al., 2015; Dorta-Gonzalez et al., 2017). Moreover, most previous studies have calculated the OA citation advantage based on the number of citations without examining the attributes of each citation, such as its field of origin.

---

[2] However, in the current landscape—characterized by the proliferation of e-journals and the implementation of policies mandating OA in some countries—it is no longer the case that only reputable authors choose to make their papers OA or that only high-quality papers are published as OA. It is also not unusual for a version of record to be made publicly available as "early access" shortly after acceptance of a manuscript. Consequently, the influence of the *early view* and *selection bias postulates* may differ between the 2000s, when these concepts were first proposed, and the present day. Notably, some recent studies have not explicitly addressed these postulates (e.g., Basson et al., 2021).

**Effects of OA from different perspectives**

The studies mentioned above focused on the OA citation advantage by examining citations among research publications, such as papers, thereby emphasizing the scholarly or academic impact of OA papers. Contrastingly, some studies have explored the social impact of OA papers by examining citations in documents other than research publications.

A notable example is research on the OA Altmetrics advantage, which investigates whether OA publications contribute to increasing individual papers' Altmetrics scores. Although many studies have reported the existence of an OA Altmetrics advantage (e.g., Cho, 2021a, 2021b; Nadavi, 2022), some have indicated that this advantage varies by field (Holmberg et al., 2020). Zong et al. (2023) investigated the relationship between paper citations and OA in policy documents and found that OA papers in the fields of library and information science led to more citations in policy documents.

Some studies focusing on the citation relationships between papers have explored aspects beyond the OA citation advantage. For instance, Huang et al. (2024) analyzed a large number of papers published between 2010 and 2019. Rather than simply examining whether making a paper OA increases the number of citations, those scholars investigated the effect of OA on the diversity of fields, regions, and institutions among the citing parties. Their findings revealed a strong correlation between OA status and an increased diversity of attributes among citing papers. Similarly, Young and Brandes' (2020) analysis of two journals found that OA papers were more likely to receive diverse citations from various fields compared to non-OA papers.

**Knowledge gap and position of this study**

This study aligns with OA citation advantage studies in that it emphasizes the scholarly impact of OA papers. However, unlike many previous studies, our focus was not on whether OA increases the total number of citations but rather on its effect on interdisciplinary citations. We were interested in the relationship between OA and interdisciplinary citations because, as mentioned earlier, this is one of the various effects OA is anticipated to produce, and to the best of our knowledge, no study has explicitly focused on it.

In this regard, this study shares common ground with studies that investigated effects beyond the traditional OA citation advantage, especially Huang et al. (2024) and Young and Brandes (2020), both of whom examined citation diversity. However, those scholars did not adopt a citation count-based metric and consequently did not provide a detailed analysis of the internal structure of disciplinary diversity, such as how many citations from which field result from making a paper in a certain field OA. Such knowledge is lacking in existing research but could be valuable to policymakers promoting OA and researchers deciding whether to make their papers OA.

This study builds on previous studies that analyzed the OA citation advantage across multiple disciplines; however, it also considered, perhaps novelly, the citing papers' disciplinary attributes. In this sense, this study falls somewhere between an OA citation advantage and an OA citation diversity study, making a novel contribution by filling the knowledge gaps found in these two types of studies.

## Methods

### Data collection and processing

*Scope and definition of the population*
This study utilized raw data from the Web of Science (WoS) core collection at the end of 2022, purchased by the National Institute of Science and Technology Policy (NISTEP). The data were included in the Science Citation Index Expanded (SCIE), the Social Sciences Citation Index, the Arts and Humanities Citation Index, and the Emerging Sources Citation Index and represented the most recent available version of WoS metadata housed at NISTEP.

We analyzed cited papers, which included gold OA and non-OA papers in the SCIE published in 2017, and citing papers, which referred to the aforementioned papers. The citing papers included those in indexes other than the SCIE. Following Dorta-González et al. (2017) and Basson et al. (2021), who, similar to our study, examined the OA citation advantage across fields using WoS data, we selected a 6-year citation window. In other words, the total number of citations in citing papers published between 2017 and 2022 was used to calculate each metric. Measuring the number of citations based on a certain citation window helps control for the *early view postulate* (Dorta-Gonzalez et al., 2017). The year 2017 was selected as the publication year of the cited papers because it was the latest publication year available within the 6-year citation window. The document type of the papers was limited to articles and reviews. This study examined the total population of papers meeting the specified criteria.

*OA type*
Although the WoS provides information on the type of OA for each paper, it reflects the paper's current status (as of data creation) and does not necessarily indicate that the paper was originally published with that type of OA. We used data from the Directory of Open Access Journals (DOAJ) to identify cited papers definitively categorized as gold OA in 2017. Specifically, we matched cited papers with DOAJ data using the journal name, ISSN, and e-ISSN as keys and identified papers published in journals whose publication year was listed as 2016 or earlier in the DOAJ. Therefore, the OA type of the cited papers in this study was gold OA, excluding hybrid OA. Contrastingly, non-OA papers were not classified under any OA type in WoS and were not listed as OA in the DOAJ.

*Field classification*

Cited and citing papers were assigned field information based on the 22 major categories in the Essential Science Indicators (ESI). The ESI assigns each journal to one of 22 field categories on a mutually exclusive basis. In other words, only one field of information was assigned to each of the cited and citing papers based on their journal source.

Additionally, only 18 fields were used for the cited papers, excluding economics and business, social sciences - general, space science, and multidisciplinary, which were among the original 22 fields. These four fields were excluded for the following reasons: First, the cited papers in this study were limited to those included in the SCIE, and the number of papers under economics and business and social sciences - general was insufficient to calculate the metrics used in the analysis. The number of gold OA papers in space science was also insufficient. Moreover, the multidisciplinary field likely contains papers from various fields, whose inclusion would have rendered the interpretation of the analysis results challenging (Basson et al., 2021). However, all 22 fields were used for the citing papers.

*JIF quartile*

As mentioned in the Literature Review, the JIF is useful for controlling *selection bias* (Niyazov et al., 2016). To calculate the metrics by JIF quartiles for each of the 18 fields, we obtained JIF data for all journals indexed in the SCIE as of 2017 (Journal Citation Reports [JCR] year 2017) from the JCR and assigned them to the cited papers by matching the journal titles.

**Metrics**

We used the proportion of the difference between the average citations of OA and non-OA papers relative to the latter as a metric of the OA citation advantage, following an approach used in cross-disciplinary studies on the OA citation advantage (Sotudeh et al., 2015; Dorta-González et al., 2017). Specifically, this metric is defined as:

$$OACA_i = \frac{OAC_i - NOAC_i}{NOAC_i} \times 100$$

Here, $OAC_i$ refers to the average citation count of OA papers in field *i*, and $NOAC_i$ refers to the average citation count of non-OA papers in field *i*. This metric, referred to as the OACA, represents *p*% more citations for OA papers compared to non-OA papers; if *p* is negative, OA papers are cited *p*% less frequently.

We also decomposed the OACA to explore whether OA increases interdisciplinary citations. Specifically, we distinguished between citations in other fields and citations in the same field and propose two metrics. The first is the IOACA, which represents the proportion of the difference between the average interdisciplinary citations of OA papers and non-OA papers to that of the latter. Interdisciplinary citation refers to papers in field *i* that are cited by papers outside of field *i*. The IOACA is defined as follows:

$$IOACA_i = \frac{IOAC_i - INOAC_i}{INOAC_i} \times 100$$

Here, $IOAC_i$ refers to the average interdisciplinary citation count of OA papers in field *i*, and $INOAC_i$ refers to the average interdisciplinary citation count of non-OA papers in field *i*. Interpretation of the IOACA values was the same as that of the OACA.

The second metric is the WOACA, which represents the proportion of the difference between the average within-discipline citations of OA papers and non-OA papers to that of the latter. Within-discipline citations refer to papers in field *i* that are cited by other papers in the same field. The WOACA is defined as follows:

$$WOACA_i = \frac{WOAC_i - WNOAC_i}{WNOAC_i} \times 100$$

Here, $WOAC_i$ refers to the average within-discipline citations of OA papers in field *i*, and $WNOAC_i$ refers to the average within-discipline citations of non-OA papers in field *i*. Interpretation of the WOACA values was the same as that of the OACA and IOACA.

Specific examples are provided to help interpret OACAs. For instance, if the OACA, IOACA, and WOACA are 29.6%, 35.6%, and 17.2% in field A, respectively, then OA papers in field A were cited 29.6% more than non-OA papers overall, 35.6% more in papers from other fields, and 17.2% more in the same field.

## Results and Discussion

**Basic information**

Table 1 presents the number of gold OA papers, non-OA papers, and the total number of papers per field. The gold OA rate varied across fields. For example, fields such as microbiology, immunology, and molecular biology exhibited high gold OA rates, whereas fields such as computer science, mathematics, and engineering showed lower rates. The gold OA rate also varied according to the JIF quartile.

**Table 1** Prevalence of gold OA by field

| Field | Abbreviation for Field | No. of gold OA papers | No. of non OA papers | No. of total papers | Percentage of gold OA papers (%) | | | | |
|---|---|---|---|---|---|---|---|---|---|
| | | | | | Q1 | Q2 | Q3 | Q4 | total |
| Agricultural Science | AGR | 7,970 | 28,367 | 47,646 | 9.49 | 6.63 | 25.60 | 37.10 | 16.73 |
| Biology & Biochemistry | BIO | 11,658 | 32,101 | 69,339 | 26.91 | 6.31 | 9.03 | 15.55 | 16.81 |
| Chemistry | CHE | 15,538 | 125,545 | 178,741 | 2.51 | 18.59 | 5.48 | 5.47 | 8.69 |
| Clinical Medicine | CLI | 45,957 | 141,961 | 282,888 | 7.59 | 25.43 | 13.80 | 15.26 | 16.25 |
| Computer Science | COM | 2,078 | 28,940 | 43,201 | 4.30 | 6.15 | 3.05 | 2.22 | 4.81 |
| Engineering | ENG | 12,905 | 113,530 | 159,042 | 3.73 | 9.72 | 10.10 | 15.22 | 8.11 |
| Environment/ Ecology | ENV | 8,444 | 35,296 | 59,057 | 4.24 | 33.01 | 6.93 | 8.25 | 14.30 |
| Geosciences | GEO | 7,771 | 28,429 | 50,450 | 12.17 | 16.00 | 26.12 | 10.33 | 15.40 |
| Immunology | IMM | 6,864 | 8,228 | 24,378 | 37.98 | 27.75 | 8.39 | 25.55 | 28.16 |
| Materials Science | MATE | 9,345 | 79,610 | 105,884 | 1.88 | 13.10 | 20.14 | 16.45 | 8.83 |
| Mathematics | MATH | 2,751 | 17,485 | 44,954 | 4.55 | 10.72 | 4.73 | 2.46 | 6.12 |
| Microbiology | MIC | 7,461 | 5,365 | 20,148 | 45.34 | 59.47 | 8.55 | 14.29 | 37.03 |
| Molecular Biology & Genetics | MOL | 10,654 | 11,760 | 47,612 | 36.75 | 18.36 | 19.54 | 16.94 | 22.38 |
| Neuroscience & Behavior | NEU | 7,581 | 21,779 | 48,994 | 8.69 | 17.51 | 22.75 | 7.38 | 15.47 |
| Pharmacology | PHA | 5,732 | 24,749 | 43,293 | 16.47 | 8.88 | 14.39 | 9.73 | 13.24 |
| Physics | PHY | 13,302 | 51,813 | 102,638 | 24.65 | 7.58 | 3.37 | 5.92 | 12.96 |
| Plant & Animal Science | PLA | 11,967 | 41,223 | 73,727 | 15.72 | 13.03 | 13.38 | 18.18 | 16.23 |
| Psychiatry/Psychology | PSY | 1,597 | 7,995 | 16,069 | 7.24 | 17.35 | 2.44 | 6.82 | 9.94 |

**OACAs by field**

Fig. 1 shows each metric calculated using the cumulative number of citations within a 6-year citation window for each field. At least one type of metric was positive in all fields except agriculture, engineering, environmental science, and materials science.

Precisely comparing this study's results with those of previous studies is difficult because of the diversity of methods used in previous studies on the OA citation advantage (Langham-Putrow et al., 2021). However, Dorta-González et al. (2017), who also used WoS data and employed the OACA to assess the OA citation advantage across disciplines with a focus on gold OA papers, identified positive OACA values in fields such as biology, chemistry, clinical medicine, immunology, and molecular biology. Although strict comparisons between this study and Dorta-González et al. (2017) are limited owing to differences in the publication years analyzed and the field classification methods, the findings appear to align to some extent. However, in our study, distinguishing between the IOACA and WOACA uncovered field-specific characteristics not captured in previous studies.

Fig. 2 plots each field with the IOACA and WOACA on the axes. The fields were divided into four groups. For Group 1 (upper right; inclusive of biology and biochemistry, geosciences, immunology, microbiology, molecular biology and genetics, neuroscience and behavior, pharmacology and toxicology, physics, plant and animal science, and psychiatry/psychology), both the IOACA and WOACA were positive. Group 2 (lower right) fields had a positive IOACA but a negative WOACA, such as chemistry, clinical medicine, and computer science. Group 3 (lower left) fields had a negative IOACA and WOACA, such as agricultural science, engineering, environment/ecology, and materials science. For Group 4 (upper left; mathematics), the WOACA was positive but the IOACA was negative.

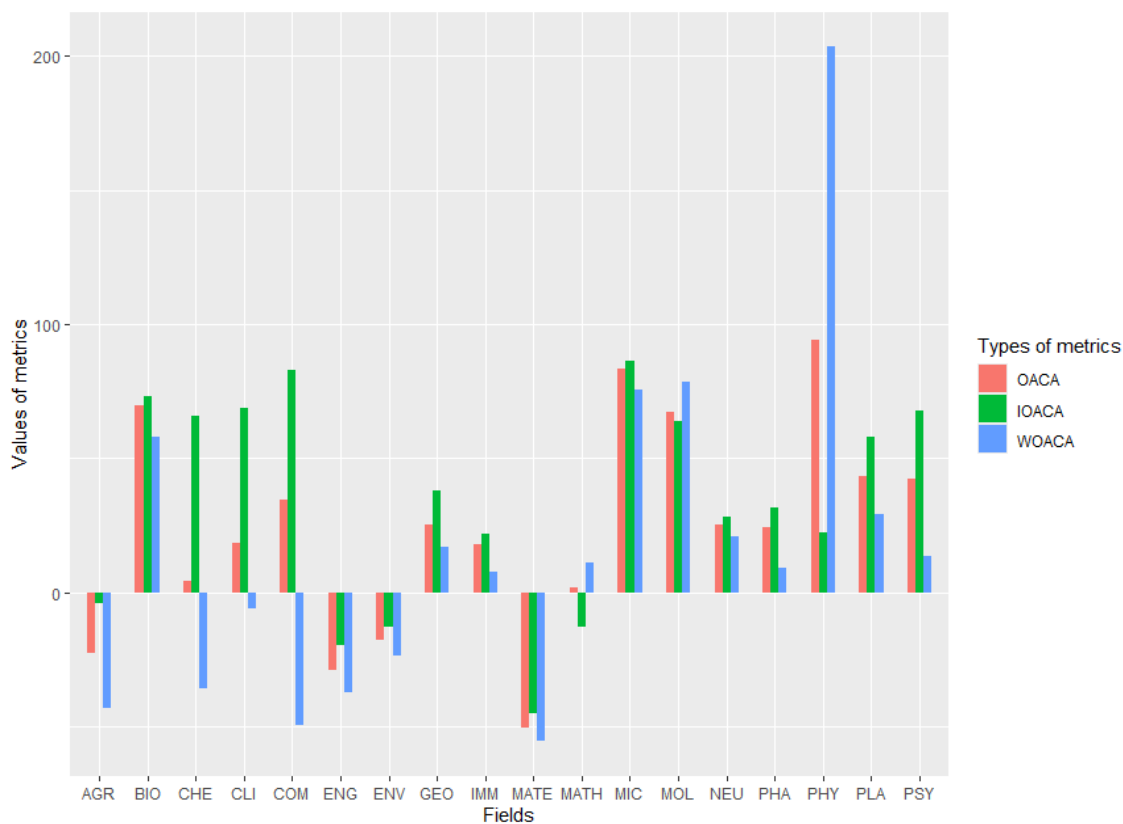

**Fig. 1** Values of OACAs by field

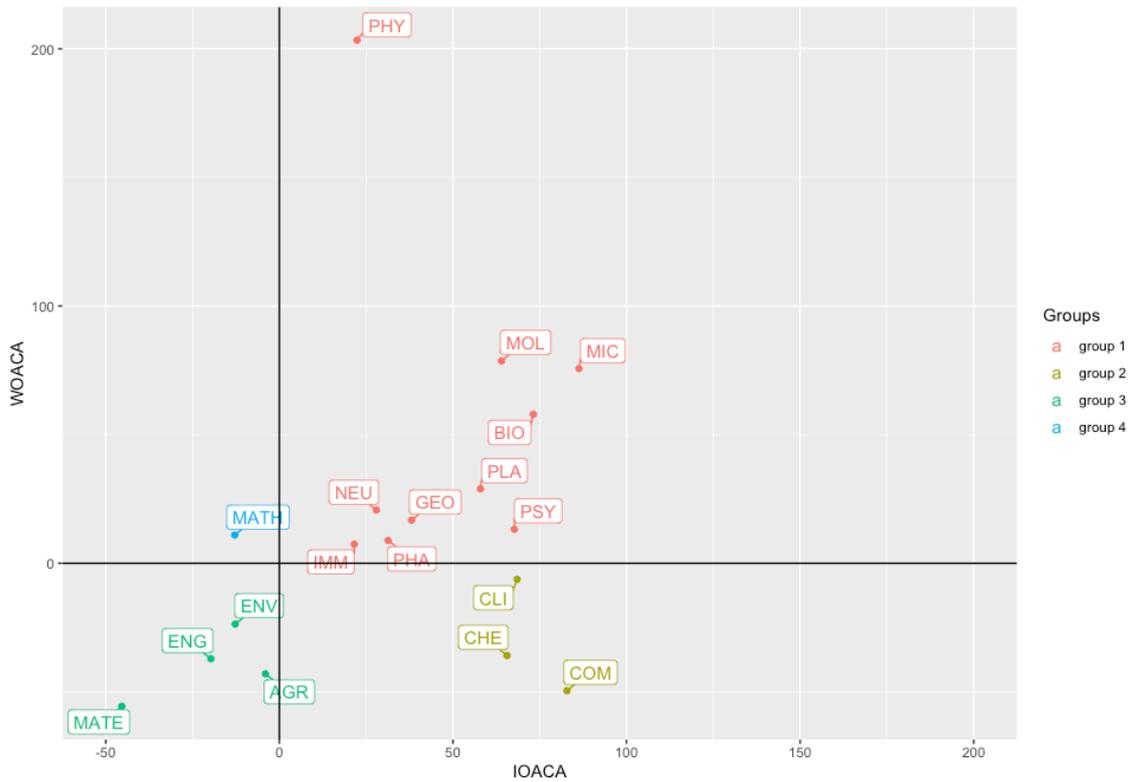

**Fig. 2** Characteristics of each field in terms of the IOACA and WOACA

Groups 1 and 2 encompass fields in which the OA citation advantage of interdisciplinary citations is evident. Group 2 is particularly unique because the OA citation advantage is seen in interdisciplinary citations only.

Figs. 1 and 2 are based on calculating each metric by summing the number of citations over a 6-year citation window. However, if a field had an unusually high number of citations in a specific year, the values of these metrics may be skewed. Therefore, confirming whether each field's characteristics have been consistent is important when calculating each metric annually. As shown in Fig. 3, some fields had negative OACAs only in the year in which the cited paper was published (2017). Nevertheless, the same general trends as those previously discussed were observed when the OACAs were calculated annually. Therefore, the remainder of this section focuses on OACAs calculated by summing the citations over a 6-year window.

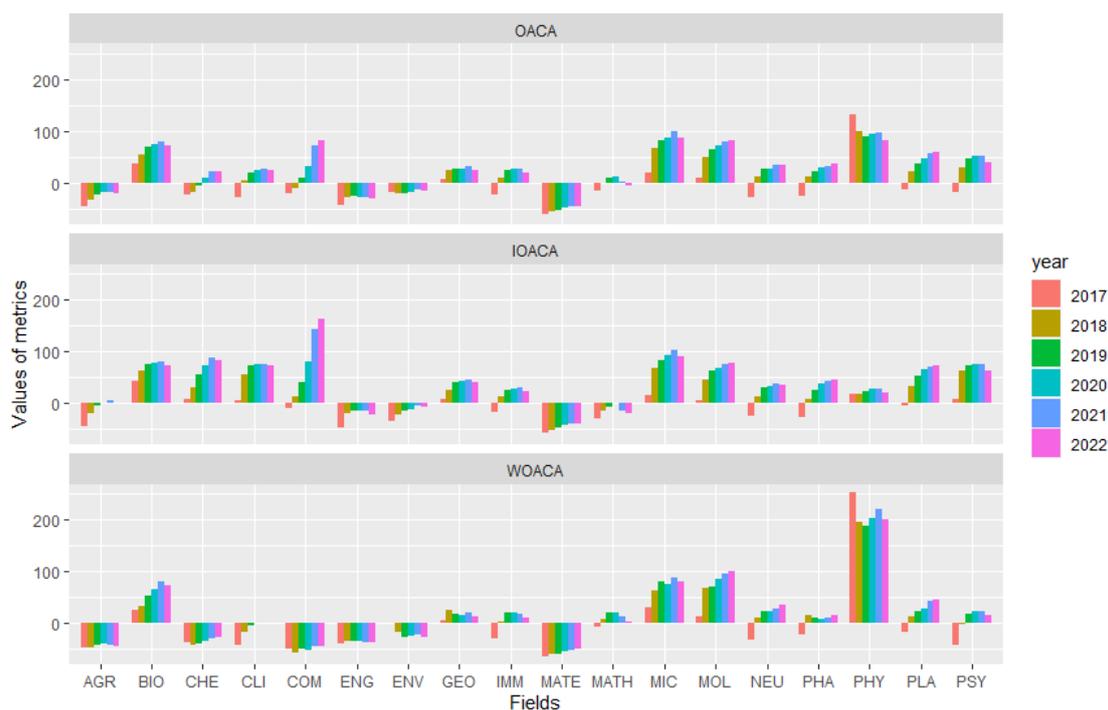

**Fig. 3** OACAs by field and year

Table 1 shows that the gold OA rate varied across JIF quartiles by field. Because journals in higher JIF quartiles generally receive more citations per paper, the OACAs for each field shown in Figs. 1 and 2 may have been influenced by the proportion of gold OA journals in higher JIF quartiles within a field. Fig. 4 presents the results of calculating OACAs using the JIF quartile for each field. Considering Table 1, Fig. 4 suggests that the relationship between gold OA rate and OACAs is not as straightforward as the higher the gold OA rate of Q1 journals, the higher the OACAs in each field.

However, in some cases, a high gold OA rate in a specific quartile along with a high value of a specific metric in the same quartile may drive a higher value of the specific metric for the field as a whole. Specifically, in physics, the high gold OA rate and WOACA in Q1 resulted in a high WOACA for the entire field. Similarly, among the fields in Group 2, chemistry had a high gold OA rate and IOACA in Q2 journals, resulting in a high IOACA for chemistry overall. In computer science, although no single quartile had an exceptionally high gold OA rate, the IOACA of Q1 journals was notably high, shaping an overall high IOACA in the field. Therefore, fields that exhibit a high OACA in a particular quartile may have been significantly influenced by specific journals in that quartile.

Although some fields, such as biology and biochemistry and clinical medicine, had high gold OA rates in specific quartiles, the OACA trends were similar across all quartiles in some fields. In these cases, the characteristics of each metric may reflect the nature of the knowledge in the field rather than having been driven by a specific journal.

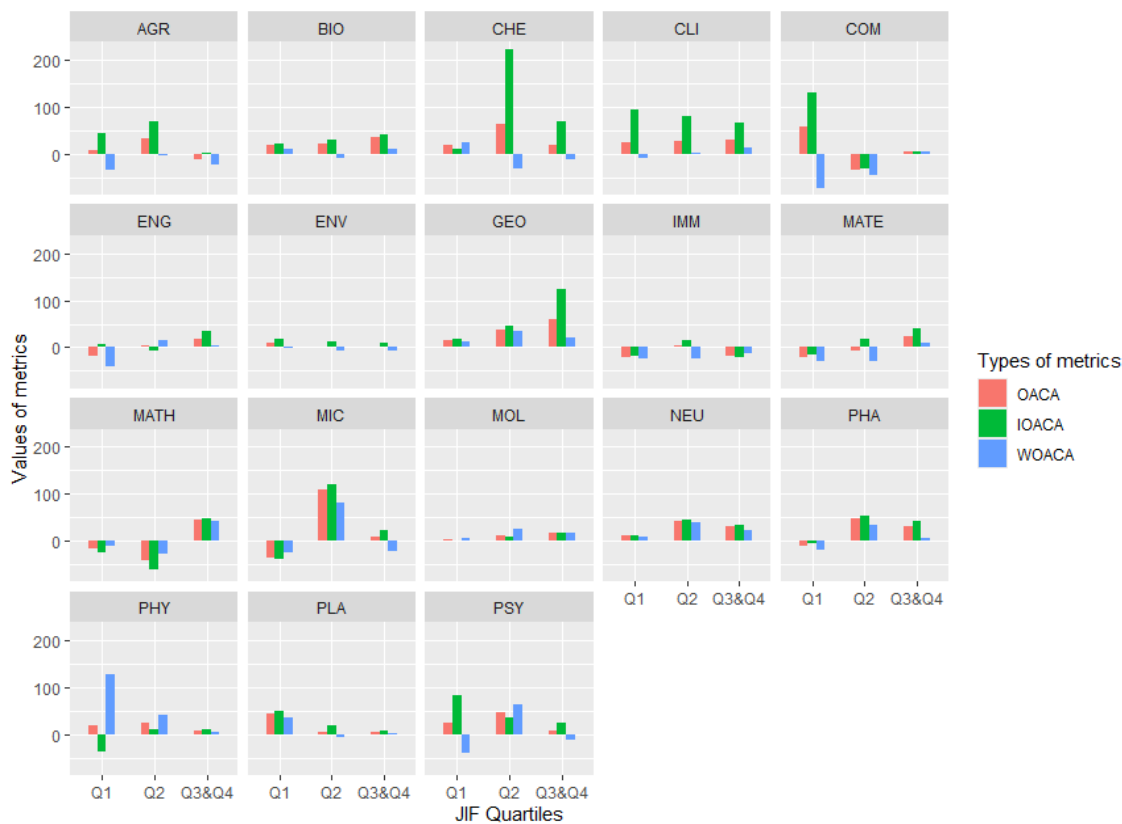

**Fig. 4** OACAs by field and JIF quartile

This study investigated the effect of OA on knowledge transfers across fields. The analysis revealed that although the Group 1 fields demonstrated the impact of OA on both interdisciplinary and within-disciplinary citations, the Group 2 fields exhibited the effect of OA solely on interdisciplinary citations. In the following sections, we explore in detail the reasons why only the IOACA values were high in the three Group 2 fields: chemistry, computer science, and clinical medicine.

**Distribution of interdisciplinary citations by journal**

As discussed in the previous section, Fig. 4 suggests that in chemistry and computer science, a higher IOACA for the whole field may be driven by certain journals in Q2 and Q1, respectively, leading to interdisciplinary citations. Therefore, we examined the number of interdisciplinary citations per journal.

Table 2 presents the number of interdisciplinary citations for papers published in Q2 gold OA chemistry journals, segmented by journal, and the corresponding percentages. The *International Journal of Molecular Sciences*, *Sensors*, and *Molecules* had a particularly high percentage of interdisciplinary citations, suggesting that they have driven the IOACA in chemistry. The JCR lists the *International Journal of Molecular Sciences* and *Molecules* under "Biochemistry & Molecular Biology" and "Chemistry, Multidisciplinary" categories, whereas *Sensors* is classified as "Chemistry, Analytical" and "Engineering,

Electrical & Electronic" categories. These classifications highlight the journals' interdisciplinary nature within chemistry.

**Table 2** Distribution of interdisciplinary citations by Q2 gold OA journal in chemistry

| Journal Title | No. of interdisciplinary citations | % |
|---|---:|---:|
| *International Journal of Molecular Sciences* | 78,633 | 36.22 |
| *Sensors* | 56,047 | 25.82 |
| *Molecules* | 43,406 | 20.00 |
| *Arabian Journal of Chemistry* | 13,847 | 6.38 |
| *Catalysts* | 8,514 | 3.92 |
| *Beilstein Journal of Organic Chemistry* | 3,555 | 1.64 |
| *Journal of Saudi Chemical Society* | 3,181 | 1.47 |
| *Frontiers in Chemistry* | 2,901 | 1.34 |
| *Chemistry Central Journal* | 2,170 | 1.00 |
| *Excli Journal* | 1,441 | 0.66 |
| *International Journal of Polymer Science* | 1,324 | 0.61 |
| *Chemistryopen* | 1,175 | 0.54 |
| *Green Chemistry Letters and Reviews* | 886 | 0.41 |
| Total | 217,080 | 100.00 |

Similar to Table 2, Table 3 shows the number of interdisciplinary citations for papers published in Q1 gold OA journals in computer science, segmented by journal, and the corresponding percentages. The *Journal of Statistical Software* and *BMC Bioinformatics* account for a large share of interdisciplinary citations, indicating that they have contributed significantly to the high IOACA in computer science. According to the JCR, the former is listed under the "Computer Science, Interdisciplinary Applications" and "Statistics & Probability" categories, whereas the latter is under the "Biochemical Research Methods", "Biotechnology & Applied Microbiology", and "Mathematical & Computational Biology" categories. These classifications emphasize the journals' interdisciplinary nature. Additionally, both journals have published papers that have received an exceptionally high number of citations, as shown in Figs. 7 and 8 (see Appendix), suggesting that the high IOACA in computer science was strongly influenced by papers that have been highly cited in other fields.

**Table 3** Distribution of interdisciplinary citations by Q1 gold OA journal in computer science

| Journal Title | No. of interdisciplinary citations | % |
|---|---:|---:|
| *Journal of Statistical Software* | 17,109 | 53.06 |
| *BMC Bioinformatics* | 12,662 | 39.27 |
| *Journal of Cheminformatics* | 2,476 | 7.68 |
| Total | 32,247 | 100.00 |

These results suggest that the high IOACA observed in Group 2 was largely driven by journals or papers in chemistry and computer science. Contrastingly, the high IOACA in clinical medicine appears to be a characteristic of the entire field. The key question, then, is: what impact do gold OA papers have on clinical medicine? In the following sections, we focus on clinical medicine and present a detailed analysis of the role of OA in this field.

**Case study of clinical medicine**

To better understand the effects of OA in clinical medicine, we examined OA and non-OA papers regarding the fields in which they were cited and the topics on which they were cited. As mentioned previously, this study focused on 18 natural science fields. To facilitate a clearer interpretation, these fields were grouped into several categories based on their proximity to clinical medicine. Specifically, we used hierarchical clustering following Ward's method to analyze the proximity of each of the 18 fields based on the number of citations from all fields, including those excluded from the present study. R and its stats packages were used to perform the analyses (R Core team, 2024). Fig. 5 shows the clustering results. Fields in the same cluster as those within clinical medicine were considered *close* fields, those in adjacent clusters were classified as *related* fields, and fields in other clusters were labeled *distant* fields. Clinical medicine is expressed as being in the *same* field.

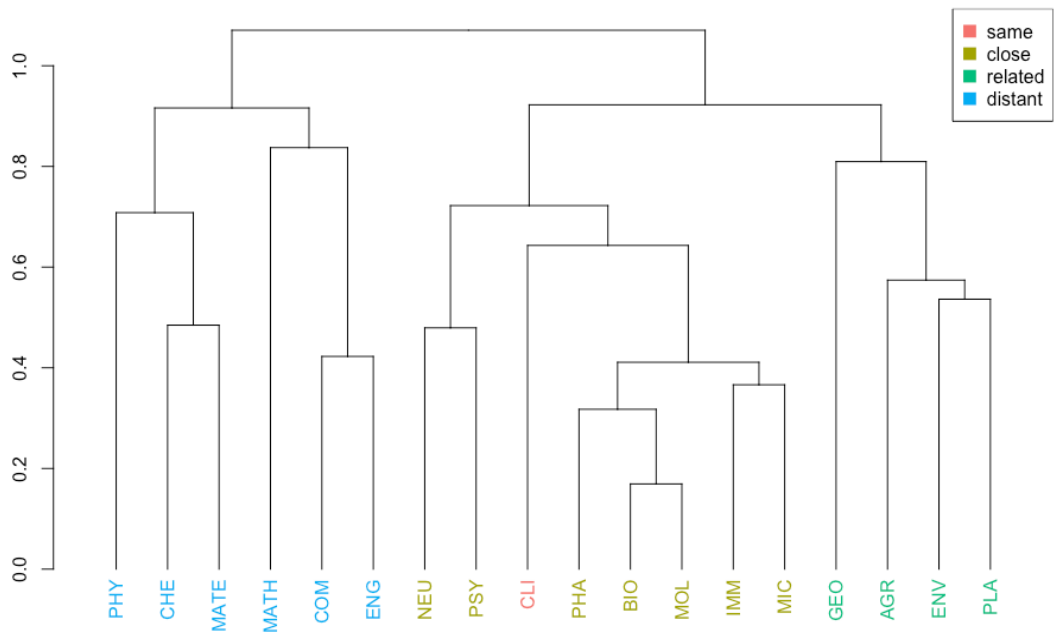

**Fig. 5** The proximity of each field to clinical medicine

To analyze each paper's topic, we used the *citation topics* provided by WoS. These topics were determined through clustering based on the citation relationships among the papers, with one topic exclusively assigned to each paper.[3] To assess whether there were differences in topics between OA and non-OA papers, we used the similarity of citation topics between the groups of papers being compared as a metric. Given the variation in the number of gold OA and non-OA papers, we employed cosine similarity, which is independent of sample size and allows for a comparison between two datasets. The cosine similarity values ranged from 1 to -1, with values closer to 1 indicating greater similarity and values closer to -1 indicating greater dissimilarity.

Fig. 6 summarizes the percentage of papers that cited clinical medicine papers by field. Table 4 shows OA and non-OA papers divided into several subsets according to the fields of the citing papers and presents the cosine similarity of the citation topics between each subset.

---

[3] Citation topics were divided into macro, meso, and micro levels. We used the meso-level citation topics for the analysis.

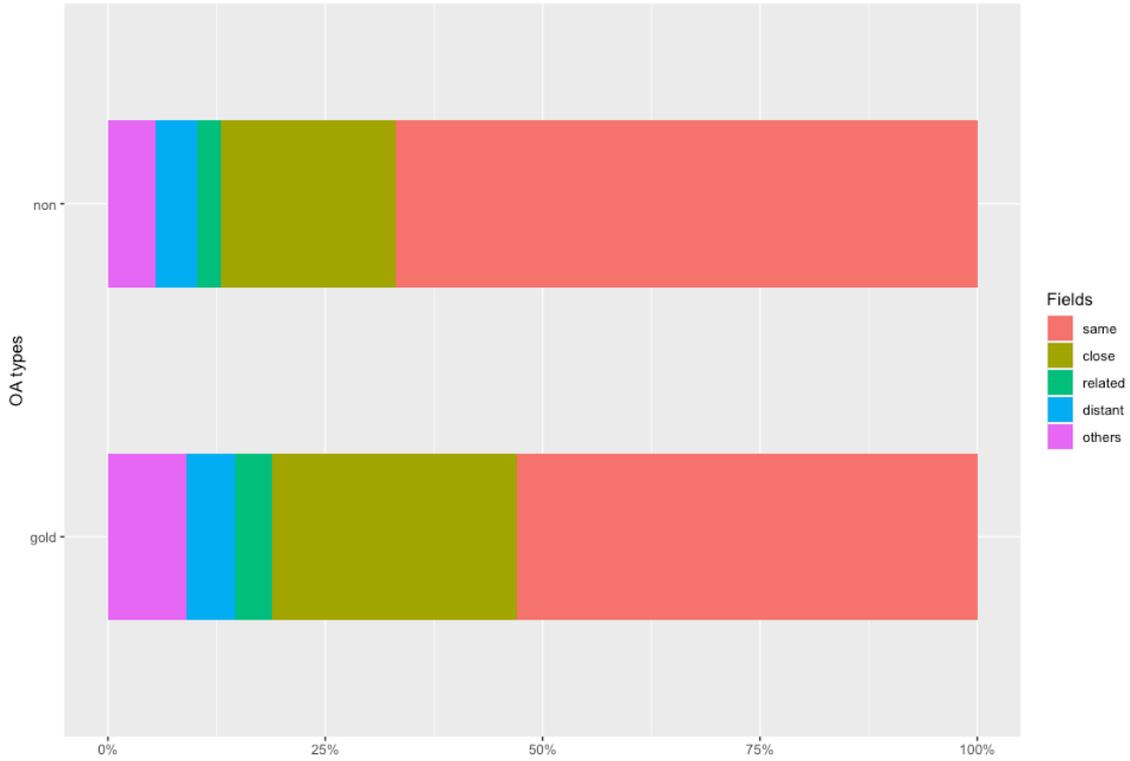

"Others" included citing papers in the four fields deemed to be outside the scope of this study's analysis, as described in the Methods section, or citing papers not assigned to an ESI category.

**Fig. 6** Percentages of citing papers in clinical medicine by field

**Table 4** Similarities among citation topics

|  | oa_all | non_all | oa_same | non_same | oa_inter | non_inter | oa_close | non_close | oa_related | non_related | oa_distant | non_distant |
|---|---|---|---|---|---|---|---|---|---|---|---|---|
| oa_all[*] | 1.00 | **0.89** | 0.97 | 0.82 | 0.96 | 0.93 | 0.90 | 0.89 | 0.70 | 0.67 | 0.82 | 0.67 |
| non_all | - | 1.00 | 0.93 | 0.98 | 0.77 | 0.91 | 0.70 | 0.84 | 0.56 | 0.63 | 0.73 | 0.78 |
| oa_same | - | - | 1.00 | **0.90** | 0.86 | 0.89 | 0.79 | 0.83 | 0.63 | 0.63 | 0.74 | 0.66 |
| non_same | - | - | - | 1.00 | 0.66 | 0.82 | 0.59 | 0.74 | 0.49 | 0.56 | 0.64 | 0.73 |
| oa_inter | - | - | - | - | 1.00 | **0.91** | 0.96 | 0.89 | 0.73 | 0.67 | 0.85 | 0.63 |
| non_inter | - | - | - | - | - | 1.00 | 0.88 | 0.96 | 0.66 | 0.71 | 0.84 | 0.80 |
| oa_close | - | - | - | - | - | - | 1.00 | **0.93** | 0.57 | 0.52 | 0.84 | 0.59 |
| non_close | - | - | - | - | - | - | - | 1.00 | 0.55 | 0.58 | 0.82 | 0.71 |
| oa_related | - | - | - | - | - | - | - | - | 1.00 | ***0.93*** | 0.53 | 0.38 |
| non_related | - | - | - | - | - | - | - | - | - | 1.00 | 0.49 | 0.45 |
| oa_distant | - | - | - | - | - | - | - | - | - | - | 1.00 | ***0.82*** |
| non_distant | - | - | - | - | - | - | - | - | - | - | - | 1.00 |

\* oa_all indicates all clinical medicine OA papers included in this study without distinguishing the field of the papers citing them. Similarly, non_all refers to all non-OA papers published in clinical medicine.

Fig. 6 shows that in clinical medicine, gold OA papers had a lower percentage of within-discipline citations (same) compared to non-OA papers and instead had more citations from particularly close fields (i.e., biology and biochemistry, immunology, microbiology, molecular biology and genetics, neuroscience and behavior, pharmacology and toxicology, and psychiatry/psychology). In this regard, Table 4 reveals that the cosine similarity between the citation topics of both OA and non-OA papers cited from close fields (i.e., between oa_close and non_close in Table 4) was as high as 0.93. This indicates that clinical medicine papers cited from close fields tended to have similar topics regardless of OA status. Furthermore, as shown in Table 4, the cosine similarity between OA and non-OA papers tended to be high when the citing paper field was consistent between the two (e.g., oa_same and no_same) and not only between oa_close and non_close papers. These findings suggest that clinical medicine OA papers were more frequently cited across different fields, not because they covered different topics than non-OA papers but because OA has improved these papers' accessibility.

Table 5 (see Appendix) lists the top 20 most cited topics by citing paper fields for both OA and non-OA papers. Topics found exclusively in citing papers in different fields included neurodegenerative diseases, phytochemicals, and inflammatory bowel diseases and infections. These three topics also appeared in both OA and non-OA papers cited in close fields but were absent in papers cited within the same field. Given that OA increased the number of citations in close fields, we can infer that papers addressing these topics are in high demand across fields and are more frequently cited because of the accessibility OA provides.

Topics common to both OA and non-OA papers that have within-discipline citations but not to those with interdisciplinary citations included cardiology - general, liver and colon cancer, and assisted ventilation. These topics' high rankings in within-discipline citations may be attributed to their specialized nature in clinical medicine, which limits their citations in other fields even when papers are available as

OA.

These findings align with those of previous studies. For example, Chen et al. (2015) explored the interdisciplinary development of biochemistry and molecular biology (BMB) over the past 100 years using WoS data and found that BMB frequently cited papers in the field of clinical medicine in particular. Consistent with Chen et al.'s (2015) findings, we observed, as depicted in Figs. 4 and 5, that BMB is closely related to clinical medicine, with clinical medicine receiving more citations from close fields due to OA.

However, according to Van Noorden (2015), who examined each field's interdisciplinarity using WoS data, clinical medicine has received relatively few citations from other disciplines. Although our study did not focus on measuring interdisciplinarity, OA may be increasing the interdisciplinarity of clinical medicine, given that the number of OA papers has increased since Van Noorden (2015), as our finding of a high IOACA in clinical medicine indicates.

**Implications and limitations**

By decomposing the OACA metric, which shows whether OA increases the overall number of citations, this study introduced a new metric, the IOACA, to measure the effect of OA on interdisciplinary citations. Our analysis revealed that OA has been shown to foster knowledge transfer in many fields. Although previous research, such as that of the OECD (2015), has suggested that OA promotes knowledge transfers across fields, our study explicitly demonstrated this effect. In this respect, this research contributes novel insights to OA citation advantage studies by focusing on citation counts. Furthermore, the findings provide policymakers, research funders, academic libraries, and individual researchers with a rationale for promoting OA.

Our findings are similar to those of Huang et al. (2024), who found that OA increases disciplinary diversity in citing papers. However, whereas Huang et al. (2024) did not specifically identify the fields more likely to be cited by papers in certain fields as a result of OA, the approach we employed in our study, which examined the effect of OA on knowledge transfers to other fields through a metric based on the citation counts of the IOACA, offers a clearer, more detailed representation of the impact of OA on interdisciplinary citation relationships. Therefore, our study explored Huang et al.'s (2024) findings in a more specific, focused manner, particularly regarding the field, and yielded insights that complement those of Huang et al. (2024).

Additionally, our results focusing on clinical medicine suggested no significant difference in topics between OA and non-OA papers. This implies that, in clinical medicine, the increased accessibility OA provides encourages citations from other fields. However, the difference in some topics between papers cited in other fields and those cited in the same field suggests that certain topics are in high demand in other fields and that OA promotes interdisciplinary citations of papers covering those topics. These findings may offer clinical medicine researchers valuable insights when deciding where and how to submit their work.

However, this study had several limitations. Firstly, it relied on the ESI classification system,

which is mutually exclusive and has a relatively small number of categories for focusing on knowledge transfers between disciplines. Although this was necessary to ensure sufficient data to calculate each metric for each field, previous studies on the OACA using the WoS have often used a more granular classification system called the subject category, making direct comparisons between this study and previous studies challenging. Secondly, because this study limited the fields of cited papers to those within the natural sciences, we were unable to capture trends in the humanities and social sciences; the effect of OA on interdisciplinary citations may also be observable in these fields. Thirdly, this study focused exclusively on gold OA without considering other forms, such as green OA; consequently, the findings may not necessarily extend to papers made available through other types of OA.

**Conclusion**

This study aimed to clarify the effects of OA on knowledge transfers across various fields. The analysis revealed that OA effectively enhances interdisciplinary citations in 13 of the 18 natural science fields examined. Furthermore, OA increases interdisciplinary citations only, especially in clinical medicine.

Regarding possible future research directions, firstly, our findings' robustness can be confirmed by conducting a similar analysis using a different field classification system or dataset. Investigating whether similar results can be obtained using Scopus or OpenAlex is particularly important given that journal coverage varies across databases. Additionally, future research could examine the effect of OA on interdisciplinary citations of papers made OA through methods other than gold OA. In particular, with green OA, where repositories and preprint servers serve as publication channels, researchers' behavior may differ from that associated with gold OA.

**Appendix**

See Fig. 7, 8 and Table 5

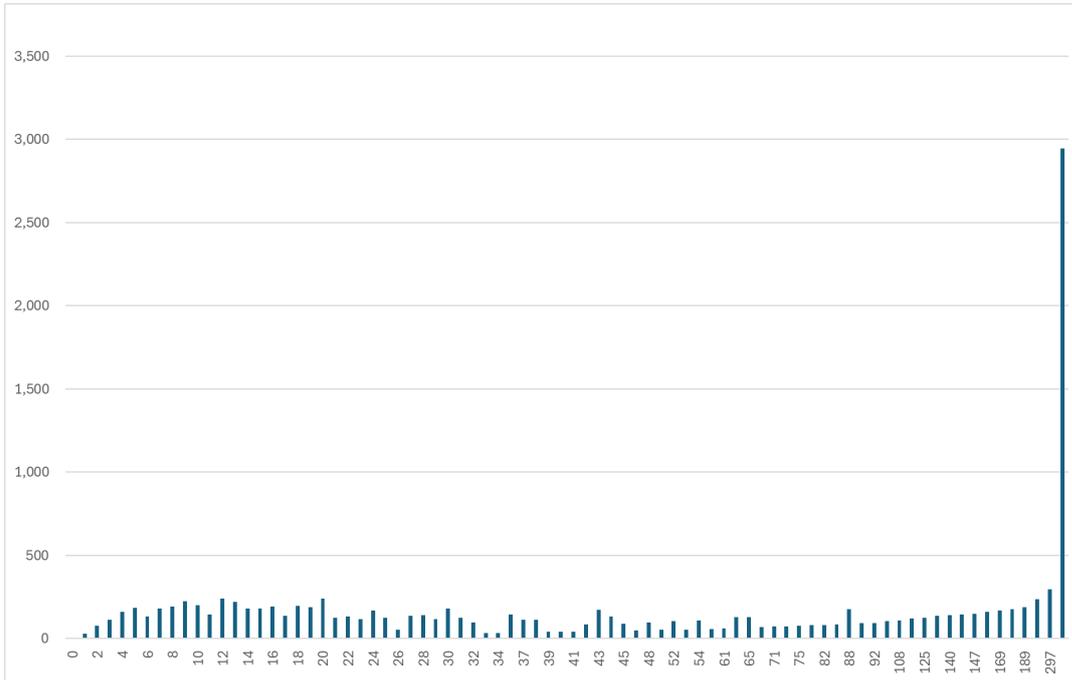

The x-axis shows the range of citations of each paper, and the y-axis shows the value of each citation multiplied by the number of papers with that value.

**Fig. 7** The citation counts of papers published in *BMC Bioinformatics*

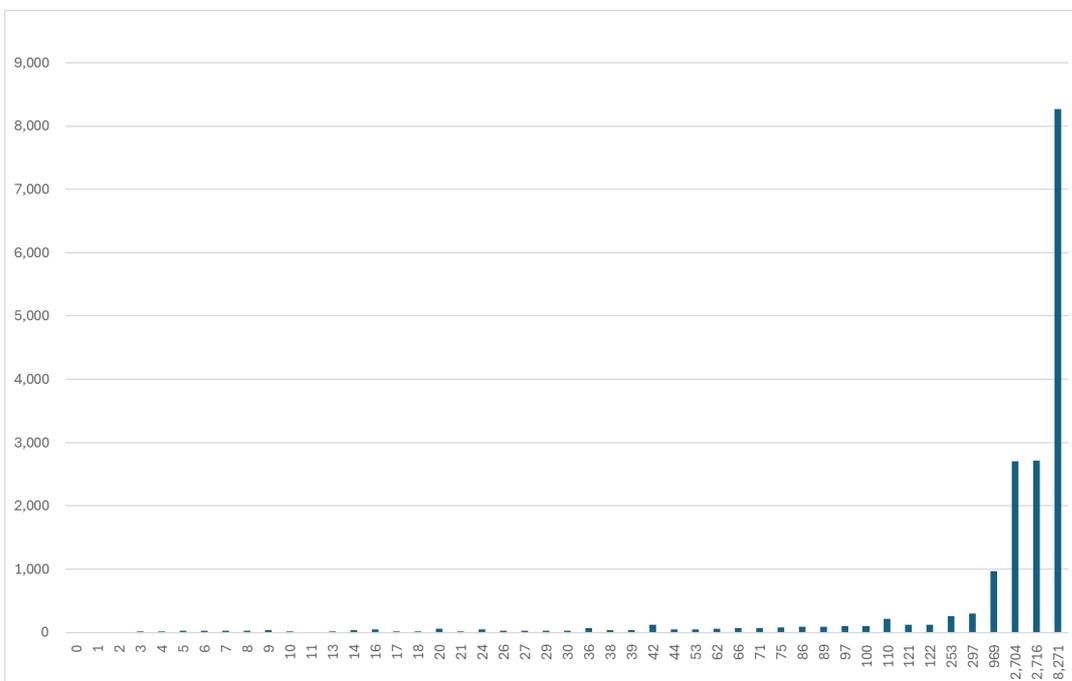

The x-axis shows the range of citations each paper has received, and the y-axis shows the value of each citation multiplied by the number of papers with that value.

**Fig. 8** The citation counts of papers published in the *Journal of Statistical Software*

**Table 5** The top 20 most cited topics by citing paper field for both OA and non-OA papers

(a) Citation topics in papers cited from different fields*

| | Gold OA | | | Non OA | | |
|---|---|---|---|---|---|---|
| Rank | Citation Topics | Citations | % | Citation Topics | Citations | % |
| 1 | Micro & Long Noncoding RNA | 16,932 | 4.96 | Immunology | 24,586 | 3.96 |
| 2 | Nutrition & Dietetics | 13,572 | 3.97 | Micro & Long Noncoding RNA | 22,051 | 3.55 |
| 3 | Immunology | 12,076 | 3.54 | Nutrition & Dietetics | 18,284 | 2.94 |
| 4 | Diabetes | 10,476 | 3.07 | Dentistry & Oral Medicine | 17,589 | 2.83 |
| 5 | Molecular & Cell Biology - Cancer, Autophagy & Apoptosis | 9,223 | 2.70 | Diabetes | 15,961 | 2.57 |
| 6 | Ophthalmology | 8,781 | 2.57 | Inflammatory Bowel Diseases & Infections | 14,419 | 2.32 |
| 7 | Neurodegenerative Diseases | 7,822 | 2.29 | Molecular & Cell Biology - Cancer, Autophagy & Apoptosis | 13,689 | 2.20 |
| 8 | Phytochemicals | 7,797 | 2.28 | Hepatitis | 11,885 | 1.91 |
| 9 | Nursing | 7,163 | 2.10 | Orthopedics | 11,689 | 1.88 |
| 10 | Molecular & Cell Biology - Cancer & Development | 6,652 | 1.95 | Nursing | 11,597 | 1.87 |
| 11 | Allergy | 6,404 | 1.88 | Rheumatology | 11,214 | 1.81 |
| 12 | Inflammatory Bowel Diseases & Infections | 5,858 | 1.72 | Obstetrics & Gynecology | 11,156 | 1.80 |
| 13 | Healthcare Policy | 5,773 | 1.69 | Ophthalmology | 10,559 | 1.70 |
| 14 | Urology & Nephrology - General | 5,376 | 1.57 | Bone Diseases | 9,767 | 1.57 |
| 15 | Parasitology - Malaria, Toxoplasmosis & Coccidiosis | 5,195 | 1.52 | Stem Cell Research | 9,545 | 1.54 |
| 16 | Obstetrics & Gynecology | 5,173 | 1.52 | Neurodegenerative Diseases | 9,127 | 1.47 |
| 17 | Antibiotics & Antimicrobials | 5,146 | 1.51 | Sports Science | 8,729 | 1.41 |
| 18 | Stem Cell Research | 5,034 | 1.47 | Allergy | 8,411 | 1.35 |
| 19 | Medical Ethics | 4,887 | 1.43 | Urology & Nephrology - General | 8,364 | 1.35 |
| 20 | Psychiatry | 4,791 | 1.40 | Lung Cancer | 8,214 | 1.32 |

* "Different fields" here refer to fields other than the same field.

(b) Citation topics in papers cited in the same field

| Gold OA | Non OA |
|---|---|

| Rank | Citation Topics | Citations | % | Citation Topics | Citations | % |
| --- | --- | --- | --- | --- | --- | --- |
| 1 | Ophthalmology | 17,109 | 4.43 | Orthopedics | 69,016 | 5.43 |
| 2 | Nutrition & Dietetics | 13,960 | 3.61 | Dentistry & Oral Medicine | 42,269 | 3.33 |
| 3 | Orthopedics | 13,686 | 3.54 | Cardiology - General | 41,188 | 3.24 |
| 4 | Diabetes | 12,325 | 3.19 | Ophthalmology | 39,923 | 3.14 |
| 5 | Immunology | 11,522 | 2.98 | Liver & Colon Cancer | 39,123 | 3.08 |
| 6 | Micro & Long Noncoding RNA | 11,470 | 2.97 | Immunology | 31,929 | 2.51 |
| 7 | Urology & Nephrology - General | 10,331 | 2.67 | Nutrition & Dietetics | 31,891 | 2.51 |
| 8 | Allergy | 10,008 | 2.59 | Nursing | 29,029 | 2.28 |
| 9 | Cardiology - General | 9,620 | 2.49 | Prostate Cancer | 27,940 | 2.20 |
| 10 | Nursing | 9,573 | 2.48 | Obstetrics & Gynecology | 27,670 | 2.18 |
| 11 | Rheumatology | 9,118 | 2.36 | Gastrointestinal & Esophageal Diseases | 25,090 | 1.97 |
| 12 | Obstetrics & Gynecology | 8,334 | 2.16 | Rheumatology | 23,414 | 1.84 |
| 13 | Liver & Colon Cancer | 7,454 | 1.93 | Assisted Ventilation | 22,911 | 1.80 |
| 14 | Assisted Ventilation | 7,252 | 1.88 | Pancreas & Gall Bladder Disorders | 22,757 | 1.79 |
| 15 | Dentistry & Oral Medicine | 7,109 | 1.84 | Diabetes | 22,578 | 1.78 |
| 16 | Breast Cancer Scanning | 6,515 | 1.69 | Breast Cancer Scanning | 21,987 | 1.73 |
| 17 | Medical Ethics | 5,812 | 1.50 | Blood Clotting | 21,935 | 1.73 |
| 18 | Palliative Care | 5,764 | 1.49 | Lung Cancer | 21,808 | 1.72 |
| 19 | Cardiology - Circulation | 5,654 | 1.46 | Cardiology - Circulation | 21,757 | 1.71 |
| 20 | Molecular & Cell Biology - Cancer, Autophagy & Apoptosis | 5,556 | 1.44 | Anesthesiology | 21,507 | 1.69 |

(c) Citation topics in papers cited in close fields

| | Gold OA | | | Non OA | | |
| --- | --- | --- | --- | --- | --- | --- |
| Rank | Citation Topics | Citations | % | Citation Topics | Citations | % |
| 1 | Micro & Long Noncoding RNA | 13,764 | 6.73 | Immunology | 19,624 | 5.21 |
| 2 | Immunology | 9,353 | 4.57 | Micro & Long Noncoding RNA | 18,197 | 4.83 |
| 3 | Molecular & Cell Biology - Cancer, Autophagy & Apoptosis | 6,991 | 3.42 | Diabetes | 10,353 | 2.75 |
| 4 | Diabetes | 6,382 | 3.12 | Molecular & Cell Biology - Cancer, Autophagy & Apoptosis | 10,304 | 2.74 |
| 5 | Neurodegenerative Diseases | 6,105 | 2.98 | Inflammatory Bowel Diseases & Infections | 9,527 | 2.53 |
| 6 | Ophthalmology | 4,944 | 2.42 | Rheumatology | 8,860 | 2.35 |
| 7 | Molecular & Cell Biology - Cancer & Development | 4,873 | 2.38 | Hepatitis | 8,685 | 2.31 |
| 8 | Nutrition & Dietetics | 4,733 | 2.31 | Neurodegenerative Diseases | 7,140 | 1.90 |
| 9 | Allergy | 4,344 | 2.12 | Stem Cell Research | 6,653 | 1.77 |
| 10 | Phytochemicals | 4,092 | 2.00 | Nutrition & Dietetics | 6,579 | 1.75 |
| 11 | Antibiotics & Antimicrobials | 3,789 | 1.85 | Allergy | 6,300 | 1.67 |
| 12 | Inflammatory Bowel Diseases & Infections | 3,581 | 1.75 | Dentistry & Oral Medicine | 5,803 | 1.54 |
| 13 | Stem Cell Research | 3,560 | 1.74 | Bone Diseases | 5,792 | 1.54 |
| 14 | Rheumatology | 3,551 | 1.74 | Strokes | 5,666 | 1.50 |
| 15 | Psychiatry | 3,521 | 1.72 | Orthopedics | 5,547 | 1.47 |
| 16 | Parasitology - Malaria, Toxoplasmosis & Coccidiosis | 3,022 | 1.48 | Obstetrics & Gynecology | 5,547 | 1.47 |
| 17 | Urology & Nephrology - General | 3,016 | 1.47 | Brain Imaging | 5,516 | 1.46 |
| 18 | Molecular & Cell Biology - Genetics | 2,745 | 1.34 | Ophthalmology | 5,450 | 1.45 |
| 19 | HIV | 2,513 | 1.23 | Lung Cancer | 5,297 | 1.41 |
| 20 | Neuroscanning | 2,445 | 1.19 | Back pain | 5,139 | 1.36 |

**Declarations**

***Competing Interests*** The authors have no competing interests to declare that are relevant to the content of this article.

***Funding*** No funding was received for conducting this study.